\documentclass{iau} 
\usepackage{graphicx}

\title[Symp. 346.~~magnetic fields in SFXTs] 
{On the origin of supergiant fast X-ray transients}

\author[S.~Hubrig et al.]   
{Swetlana~Hubrig$^1$,
Lara~Sidoli$^2$,
Konstantin~A.~Postnov$^3$,
Markus~Sch{\"o}ller$^4$,
Alexander~F.~Kholtygin$^5$,
 \and Silva~P.~J{\"a}rvinen$^1$}

\affiliation{$^1$Leibniz-Institut f\"ur Astrophysik Potsdam (AIP),
An der Sternwarte~16, 14482~Potsdam, Germany, email: {\tt shubrig@aip.de} \\[\affilskip]
$^2$INAF, Istituto di Astrofisica Spaziale e Fisica Cosmica, Via E.~Bassini~15, 20133~Milano, Italy\\
$^3$Sternberg Astronomical Institute, Moscow M.V. Lomonosov State University, 119234~Moscow, Russia\\
$^4$European Southern Observatory, Karl-Schwarzschild-Str.~2, 85748~Garching, Germany\\
$^5$Saint-Petersburg State University, Universitetskij pr.~28, 198504~Saint-Petersburg, Russia
}

\pubyear{2018}
\volume{346}  
\jname{High-mass X-ray binaries:
illuminating the passage from massive binaries
to merging compact objects}
\editors{A.C. Editor, B.D. Editor \& C.E. Editor, eds.}
\begin{document}

\maketitle

\begin{abstract}
A fraction of high-mass X-ray binaries are supergiant fast X-ray transients.
These systems have on average low X-ray luminosities, but display short flares 
during which their X-ray luminosity rises by a few orders of magnitude. The 
leading model for the physics governing this X-ray behaviour suggests that the 
winds of the donor OB supergiants are magnetized. In agreement with this model,
the first spectropolarimetric observations of the SFXT IGR J11215-5952 using the 
FORS\,2 instrument at the Very Large Telescope indicate the presence of a kG longitudinal 
magnetic field. Based on these results, it seems possible that the key 
difference between supergiant fast X-ray transients and other high-mass X-ray 
binaries are the properties of the supergiant's stellar wind and the physics of 
the wind's interaction with the neutron star magnetosphere.
\end{abstract}

\firstsection 

\section{Introduction}

Among the bright X-ray sources in the sky, a significant number contain a compact 
object (either a neutron star or a
black hole) accreting from the wind of a companion star with a mass above $10\,M_\odot$.
Such systems are 
called high-mass X-ray binaries (HMXBs). They are young (several
dozen million years old) and can be formed when one of the initial binary members 
loses a significant part of its mass through stellar wind or mass transfer before a first
supernova explosion occurs (van den Heuvel \& Heise 1972). 

Supergiant Fast X-ray Transients (SFXTs) are a subclass of HMXBs associated with early-type
supergiant companions, and characterized by sporadic, short and bright X--ray flares
reaching peak luminosities of 10$^{36}$--10$^{37}$\,erg\,s$^{-1}$ and typical energies released in 
bright flares of about 10$^{38}$--10$^{40}$\,erg -- see the review by Sidoli (2017) for more details. 
Their X-ray spectra in outburst 
are very similar to accreting pulsars in HMXBs. In fact, half of them have measured neutron
star spin periods similar to those observed from persistent HMXBs (Shakura et al.\ 2015, 
Martinez-Nunez et al.\ 2017).
The physical mechanism driving their transient behavior,
probably related to the accretion of matter from the supergiant wind by the compact object,
has been discussed by several authors and is still
a matter of debate. The leading model for the existence of SFXTs
invokes their different wind properties and magnetic field strengths  
that lead to distinctive accretion regimes
(Shakura et al.\ 2012,
Postnov et al.\ 2015,  
Shakura et al.\ 2015, Shakura \& Postnov 2017).
The SFXTs' behaviour can be explained 
by sporadic capture of magnetized stellar wind. 
The effect of the magnetic field carried by the stellar wind is twofold: first, it may
trigger rapid mass entry to the magnetosphere via magnetic reconnection in the 
magnetopause (a phenomenon that is well known in the dayside of Earth's
magnetosphere), and
secondly, the magnetized parts of the wind (magnetized clumps with a tangent 
magnetic field) have a lower velocity than the non magnetised parts (or the parts carrying the
radial field; Shakura et al.\ 2015).
The model predicts that a magnetized clump 
of stellar wind with a magnetic field strength of a few 10\,G triggers  
sporadic reconnection, allows accretion, and results in an X-ray 
flare. Typically, the  neutron star orbital separation is a few $R_{\ast,{\rm RSG}}$. 
Thus, the expected required magnetic field on the stellar surface is of the order of 
$100-1000$\,G.

\section{Magnetic field measurements}

\begin{figure}
\centering
\includegraphics[width=0.68\textwidth]{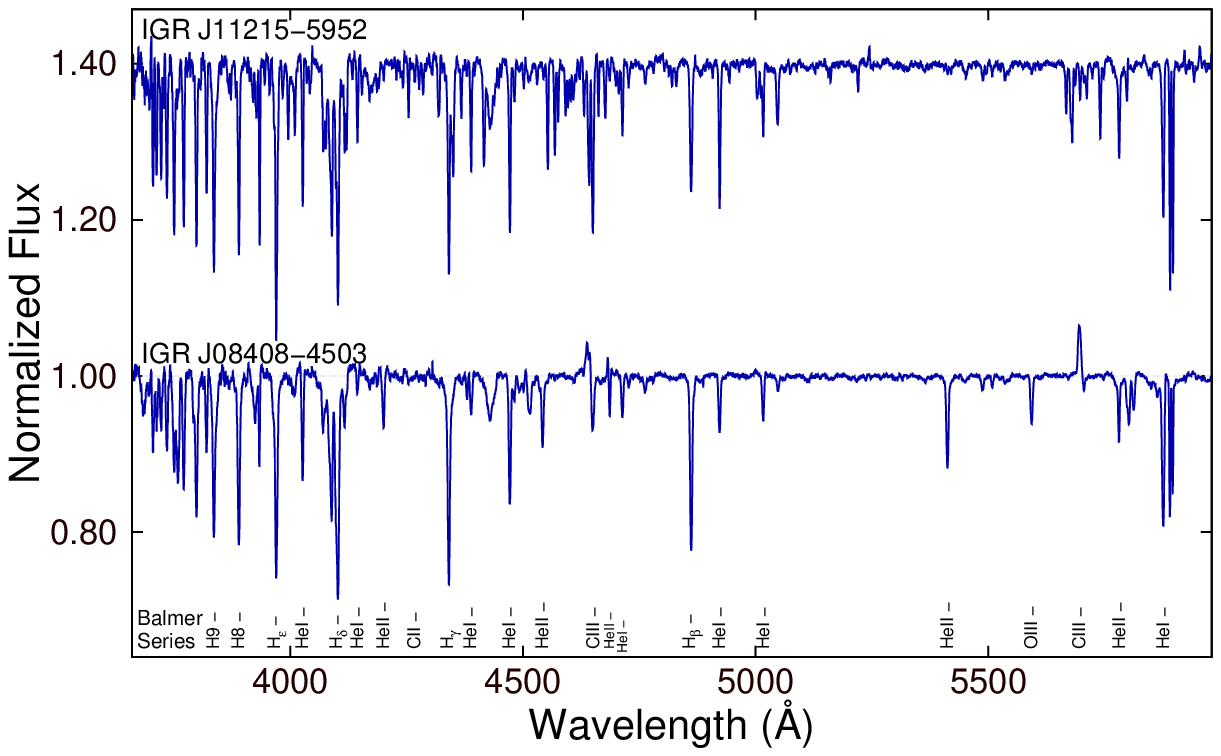}
\caption{ 
Normalised FORS\,2 Stokes~$I$
spectra of IGR\,J08408$-$4503 and IGR\,J11215$-$5952.
Well known spectral lines are indicated at the bottom.
The spectrum of IGR\,J11215$-$5952 was vertically offset by 0.4 for clarity.
}
\label{fig1}
\end{figure}

To investigate the magnetic nature of SFXTs, we recently observed the two optically brightest targets, 
IGR\,J08408-4503 ($P_{\rm orb}=9.5$\,d) and IGR\,J11215-5952 ($P_{\rm orb}=165$\,d), 
using the FOcal Reducer low dispersion 
Spectrograph (FORS\,2; Appenzeller et al.\ 1998) mounted on the 8\,m Antu telescope of 
the Very Large Telescope in spectropolarimetric mode.
No significant magnetic field was measured in the spectra of IGR\,J08408-4503,
with the highest value $\left<B_{\rm z}\right>_{\rm hyd}=-184\pm97$\,G at a significance level
of 1.9$\sigma$.
On the other hand, a definite magnetic field detection was achieved for IGR\,J11215-5952 in 2016 December
with $\left<B_{\rm z}\right>_{\rm hyd}=416\pm110$\,G with a significance 
at the 3.8$\sigma$ level.
The measurement obtained in 2016 May yielded $\left<B_{\rm z}\right>_{\rm hyd}=-978\pm308$\,G at 
a significance level of 3.2$\sigma$ (Hubrig et al.\ 2018).
The  spectral  appearance  of IGR\,J08408$-$4503 and IGR\,J11215$-$5952 in the FORS\,2 spectra
is presented in Fig.~\ref{fig1}.

\begin{figure}
\centering
\includegraphics[width=0.68\textwidth]{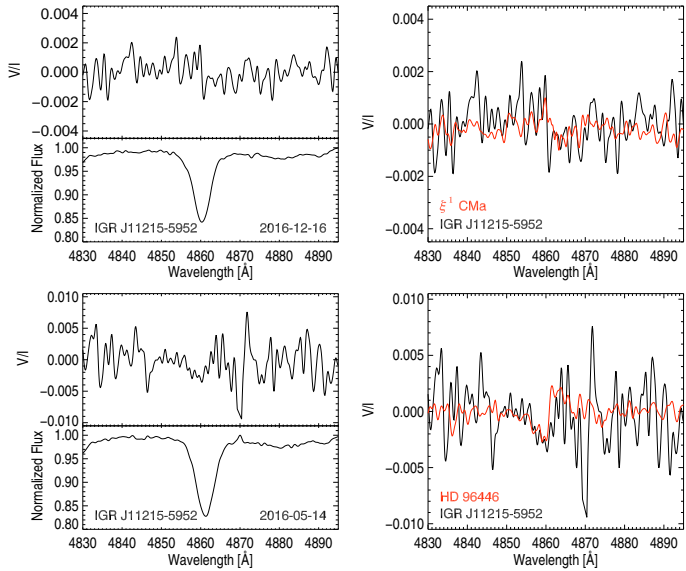}
\caption{ 
{\it Left panel}: Stokes~$V$  and Stokes~$I$ spectra of IGR\,J11215$-$5952 in the spectral region 
around the H$\beta$ line at two different epochs.
{\it Right panel}: Stokes~$V$ spectra of IGR\,J11215$-$5952 overplotted with the Stokes~$V$ spectra
of the two well-known magnetic early-B type stars  $\xi^1$\,CMa ($\left< B_{\rm z}\right>_{\rm hydr}=360\pm49
$\,G)
and HD\,96446 ($\left< B_{\rm z}\right>_{\rm hydr}=-1590\pm74$\,G) for best visibility of the Zeeman features.
}
\label{fig:zf}
\end{figure}

In Fig.~\ref{fig:zf}, we present Stokes~$V$ spectra of IGR\,J11215$-$5952 obtained on these two nights 
in the spectral region around the H$\beta$ line. For best visibility of the Zeeman features, we overplot 
the Stokes~$V$ spectra of IGR\,J11215$-$5952 with the Stokes~$V$ spectra
of the two well-known magnetic early B-type stars HD\,96446 and $\xi^1$\,CMa.

Regarding the significance of the magnetic field detections in massive stars at 
significance levels around 3$\sigma$, we note that the two clearly magnetic Of?p stars
HD\,148937 and CPD~$-$28$^\circ$5104 have been
for the first time detected as magnetic in our FORS\,2 observations at significance 
levels of 3.1$\sigma$ and 3.2$\sigma$, 
respectively (Hubrig et al.\ 2008, Hubrig et al.\ 2011).
The detection of a magnetic field in IGR\,J11215$-$5952 at significance levels of 3.2$\sigma$ and
3.8$\sigma$ indicates that this target likely possesses a kG magnetic field.
 Although no significant magnetic field was 
measured in the spectra of IGR\,J08408-4503, due to the presence of distinct Zeeman features
in its spectra  it appears
still valuable to obtain additional spectropolarimetric observations 
on other epochs corresponding to different orbital phases.

\section{Discussion}

Our spectropolarimetric observations of IGR\,J11215$-$5952 
revealed the presence of a magnetic field on two occasions.
This target is the only SFXT where strictly periodic X-ray outbursts have been observed, 
repeating every 164.6\,d (Sidoli et al.\ 2006, Sidoli et al.\ 2007, Romano et al.\ 2009).
To explain these short periodic outbursts, Sidoli et al.\ (2007) proposed that they are triggered 
by the passage 
of the neutron star inside an equatorial enhancement of the outflowing supergiant wind, focussed on a plane 
inclined with respect to the orbit. 
This  configuration of the line-driven stellar wind might be magnetically channeled (ud-Doula \& Owocki 
2002).
The effectiveness of the stellar magnetic field in focussing the wind is indicated by the
wind magnetic confinement parameter $\eta$
defined as $\eta= B_{*} ^2 R_{*}^2$ / ($\dot M_{\rm w}\upsilon_{\infty}$),
where B$_{*}$ is the strength of the magnetic field at the surface of the supergiant, R$_{*}$ is the stellar radius
(R$_{*}$=40~R$_\odot$), $\upsilon_{\infty}$ is the wind terminal velocity ($\upsilon_{\infty}=1200~km~s^{-1}$)
and $\dot{M}_{\rm w}$ is the wind mass loss rate ($\dot M_{\rm w}=10^{-6}$\,M$_\odot$\,yr$^{-1}$; 
Lorenzo et al.\ 2014).
Adopting B$_{*} \ge$ 0.7\,kG at the magnetic equator, we estimate $\eta \ge$ 500, 
implying a wind 
confinement, up to an Alfv\'en radius $R_A=\eta^{1/4}R_{*} \ge 4.73\,R_{*}$  (ud-Doula \& Owocki 2002). This 
radial distance is 
compatible with the orbital separation at periastron in IGR\,J11215$-$5952,
where the orbital eccentricity is high ($e>0.8$; Lorenzo et al.\ 2014). 
The measured magnetic field strength in IGR\,J11215$-$5952 reported here for the first time is high enough to channel 
the stellar wind on the magnetic equator, supporting the scenario proposed by Sidoli et al.\ (2007) 
to explain the 
short periodic outbursts in this SFXT. 

Because of the faintness of SFXTs -- most of them have a visual magnitude $m_V \ge 12$, up to  $m_V \ge 31$ 
(Sidoli 2017, Persi et al.\ 2015) --
no high-resolution spectropolarimetric observations were carried out for these objects so far and the presented FORS\,2 
observations are the first to explore the magnetic nature of the optical counterparts.
Future spectropolarimetric observations 
of a representative sample of SFXTs are urgently needed to be able to draw solid conclusions about
the role of magnetic fields in the wind accretion process and to 
constrain the conditions that enable the presence of magnetic
fields in massive binary systems.

\end{document}